\documentclass[superscriptaddress,showpacs,aps,twocolumn]{revtex4}
\usepackage{amssymb}
\usepackage{amsmath}
\usepackage{graphicx}
\usepackage{color}
\usepackage{amsfonts}
\usepackage{multirow}

\begin{document}

\title{Phonon contribution in grazing-incidence fast atom diffraction from
insulator surfaces}
\author{L. Frisco}
\affiliation{Instituto de Astronom\'{\i}a y F\'{\i}sica del Espacio (UBA-CONICET).
Casilla de Correo 67, Sucursal 28, (C1428EGA) Buenos Aires, Argentina.}
\author{M.S. Gravielle\thanks{%
Author to whom correspondence should be addressed.\newline
Electronic address: msilvia@iafe.uba.ar}}
\affiliation{Instituto de Astronom\'{\i}a y F\'{\i}sica del Espacio (UBA-CONICET).
Casilla de Correo 67, Sucursal 28, (C1428EGA) Buenos Aires, Argentina.}
\date{\today }

\begin{abstract}
We study the effect of crystal lattice vibrations on grazing-incidence fast
atom diffraction (GIFAD) from insulator surfaces. To describe the phonon
contribution to GIFAD we introduce a semi-quantum method, named Phonon-
Surface Initial Value Representation (P-SIVR), which represents the surface
with a harmonic crystal model, while the scattering process is described by
means of the Surface Initial Value Representation approach, including phonon
excitations. Expressions for the partial scattering probabilities involving
zero- and one- phonon exchange are derived. In particular, the P-SIVR
approach for zero-phonon scattering is applied to study the influence of
thermal lattice vibrations on GIFAD patterns for Ne/LiF(001) at room
temperature. It is found that the thermal lattice fluctuations introduce a
polar-angle spread into the projectile distributions, which \ can affect the
relative intensities of the interference maxima, even giving rise to
interference sub-patterns depending on the incidence conditions. Present
results are in agreement with the available experiments.
\end{abstract}

\maketitle

\section{Introduction}

Like in any interference phenomenon, in grazing-incidence fast atom
diffraction (GIFAD or FAD) from ordered surfaces, the observation of
interference
patterns depends on the coherence conditions \cite%
{Lienemann2011,Seifert2015,Gravielle2015}. In this regard, since the early
reports of GIFAD \cite{Schuller2007,Rousseau2007} thermal lattice vibrations
were suspected of deteriorating the coherence, making the observation of
interference structures completely unexpected \cite{Aigner2008,Manson2008}.
This was based on the fact that in typical GIFAD experiments the de Broglie
wavelengths of the projectiles are much smaller than the mean thermal
fluctuations of the surface atoms, which would suggest a strong coherence
loss. However, over the last decade GIFAD \ was observed for a wide variety
of materials at room temperature \cite%
{Bundaleski2008,Schuller2009b,Seifert2010,Atkinson2014,Seifert2012,Seifert2013,Zugarramurdi2015,Momeni2018}%
, indicating that the quantum interference prevails over the decoherence
mechanisms. Moreover, GIFAD patterns were found to be extremely sensitive to
the projectile-surface interaction, allowing the determination of surface
parameters smaller than the thermal vibration amplitudes, like rumpling \cite%
{Schuller2010} or corrugation \cite{Debiossac2016prb} distances.

From the theoretical point of view, in spite of the above mentioned
features, most of the GIFAD models \cite%
{Schuller2012,Debiossac2014,Gravielle2014,delCueto2017} consider an ideal
and static crystal surface, with atoms or ions at rest at their equilibrium
positions. On the other hand, few articles deal with the decoherence
introduced in GIFAD by lattice vibrations \cite%
{Manson2008,Aigner2008,Schuller2010,Roncin2017,Roncin2018}, so this issue
represents a problem not fully understood yet.

In this paper we study the effect of lattice vibrations, i.e., phonons, on
GIFAD distributions from insulator surfaces. This kind of surfaces is a good
candidate to investigate the partial decoherence introduced by phonons
because the presence of a wide band-gap strongly suppresses the electronic
excitations of the target, causing the main source of decoherence to come
from the vibrational movements of the surface atoms \cite%
{Taleb2017,Schram2018}.

With a view to describe the collision with a realistic crystal that enables
phonon transitions, we develop a semi-quantum method, named Phonon-Surface
Initial Value Representation (P-SIVR) approximation. It is based on the
previous SIVR approach for elastic scattering from a rigid surface \cite%
{Gravielle2014}, which was successfully employed to describe experimental
GIFAD patterns for different collision systems \cite%
{Bocan2016,Gravielle2015,Frisco2018,Bocan2018}. The basic idea of the P-SIVR
method is to incorporate a quantum representation of the surface, given by
the harmonic crystal model \cite{Ashcroft}, making possible the description
of the phonon effects involved in the GIFAD process.

Within the P-SIVR approximation, the scattering probability can be expressed
as a series on the number $n$ of phonons emitted or absorbed during the
collision. Each term of the series, named here as P$n$-SIVR probability, is
associated with the grazing scattering involving the exchange of $n$
phonons. Mathematical formulas for the\ P$n$-SIVR probabilities
corresponding to $n=0$ and $n=1$ exchanged phonons are presented.

In this work the P0-SIVR approach for zero-phonon scattering is applied to
study the influence of thermal lattice vibrations on GIFAD patterns for
Ne/LiF(001) at room temperature. With the aim of examining the thermal
contribution, P0-SIVR results for different incidence conditions are
compared with values derived from the SIVR approach for the static surface.
In all the considered cases, it is found that the thermal lattice vibrations
contribute to the polar-angle spread of the projectile distributions, in
accord with previous predictions \cite{Manson2008}. Furthermore, depending
on the incidence conditions, such thermal fluctuations can affect the
relative intensity of the interference peaks, even introducing an
interference sub-pattern, as it is observed at a high normal energy. We show
that present P0-SIVR results are in very good agreement with the available
experimental data \cite{Gravielle2011}. In addition, results from an
incoherent model to include thermal lattice vibrations are analyzed.

The article is organized as follows. The P-SIVR approach is summarized in
Sec. II, while details about its derivation are given in the Appendix.
Results are presented and discussed in Sec. III and in Sec. IV we outline
our conclusions. Atomic units (a.u.) are used unless otherwise stated.

\section{Theoretical model}

The P-SIVR approximation can be considered as a natural extension of the
SIVR approach \cite{Gravielle2014} to incorporate phonon effects into the
GIFAD description. Summarizing, within the P-SIVR approximation the
atom-surface scattering probability corresponding to the transition $\mathbf{%
K}_{i}\rightarrow \mathbf{K}_{f}$, with $\mathbf{K}_{i}$ ($\mathbf{K}_{f}$)
being the initial (final) projectile momentum, is evaluated by adding the
partial contributions coming from the different initial and final crystal
states, which are derived from a quantum harmonic-crystal model \cite%
{Ashcroft}. The result is then expanded in terms of the number $n$ of
phonons that are exchanged with the crystal, giving rise to a series of P$n$%
-SIVR probabilities associated with $n$- phonon scattering. In this Section
we present mathematical expressions of the P$n$-SIVR probability for $n=0$
and $n=1$, while the general formula for a given value of $n$, as well as
the steps involved in its derivation, are given in the Appendix.

\medskip The \ P0-SIVR probability for the\ transition $\mathbf{K}%
_{i}\rightarrow \mathbf{K}_{f}$ \ without phonon exchange, which corresponds
to the elastic scattering with $K_{f}=K_{i}$, can be expressed as
\begin{equation}
\frac{dP_{0}}{d\mathbf{K}_{f}}=\left\vert \mathcal{A}_{n_{ph}=0}\right\vert
^{2},  \label{prob-P0-sivr}
\end{equation}%
while the P1-SIVR probability for one-phonon scattering reads
\begin{eqnarray}
\frac{dP_{1}}{d\mathbf{K}_{f}} &=&\sum\limits_{\mathbf{k},l}\left[ \frac{%
N_{l}(\mathbf{k})}{\omega _{l}(\mathbf{k})}\left\vert \mathcal{A}%
_{n_{ph}=-1}(\mathbf{k},l)\right\vert ^{2}\right.   \notag \\
&&\left. +\frac{N_{l}(\mathbf{k})+1}{\omega _{l}(\mathbf{k})}\left\vert
\mathcal{A}_{n_{ph}=+1}(\mathbf{k},l)\right\vert ^{2}\right] ,
\label{prob-P1-sivr}
\end{eqnarray}%
where $\mathcal{A}_{n_{ph}}$ is the effective transition amplitude for
scattering involving $n_{ph}=\pm $ $n$ phonons emitted ($n_{ph}=-n$) or
absorbed ($n_{ph}=+n$) by the crystal. In Eq. (\ref{prob-P1-sivr}) the sum
runs over all the normal modes of the crystal, with $\omega _{l}(\mathbf{k})$
being the phonon frequency in the branch $l$, with the wave vector $\mathbf{k%
}$. The factor $N_{l}(\mathbf{k})=\left( \exp \left[ \omega _{l}(\mathbf{k}%
)/\left( k_{B}T\right) \right] -1\right) ^{-1}$ \ is the Bose-Einstein
occupation function for $(\mathbf{k},l)$ phonons in a crystal target at
temperature $T$, with\ $k_{B}$ being the Boltzmann constant.

In Eqs. (\ref{prob-P0-sivr}) and (\ref{prob-P1-sivr}), the effective
transition amplitude $\mathcal{A}_{n_{ph}}(\mathbf{k},l)$ reads
\begin{eqnarray}
\mathcal{A}_{n_{ph}}(\mathbf{k},l) &=&\int d\mathbf{R}_{o}\ f(\mathbf{R}%
_{o})\int d\mathbf{K}_{o}\ g(\mathbf{K}_{o})  \notag \\
&&\times \ \ \int d\underline{\mathbf{u}}_{o}\ a_{n_{ph}}^{(\mathbf{k},l)}(%
\mathbf{R}_{o},\mathbf{K}_{o},\underline{\mathbf{u}}_{o}),\text{ \ \ }
\notag \\
&&\text{\ \ \ }  \label{APn}
\end{eqnarray}%
where the function $f$ ($g$) describes the position (momentum) profile of
the incident wave packet and
\begin{eqnarray}
a_{n_{ph}}^{(\mathbf{k},l)}(\mathbf{R}_{o},\mathbf{K}_{o},\underline{\mathbf{%
u}}_{o}) &=&\int\limits_{0}^{+\infty }dt\ \left\vert J_{P}(t)\right\vert
^{1/2}e^{i\nu _{t}\pi /2}\ \mathcal{V}_{n}^{^{(\mathbf{k},l)}}(\mathbf{R}%
_{t})  \notag \\
&&\times \exp \left[ i\left( \varphi _{t}-\mathbf{Q}\cdot \mathbf{R}%
_{o}+n_{ph}\omega _{l}(\mathbf{k})t\right) \right]  \notag \\
&&\quad  \label{bPn}
\end{eqnarray}%
represents the partial amplitude associated with the classical projectile
trajectory $\mathbf{R}_{t}\equiv \mathbf{R}_{t}(\mathbf{R}_{o},\mathbf{K}%
_{o},\underline{\mathbf{u}}_{o})$, which starts at the position $\mathbf{R}%
_{o}$ with momentum $\mathbf{K}_{o}$ and it is determined by the spatial
configuration $\underline{\mathbf{u}}_{o}$ of the crystal at the initial
time $t=0$. That is, the underlined vector $\underline{\mathbf{u}}_{o}$
denotes the $3N$-dimension vector associated with the spatial deviations of
the $N$ ions contained in the crystal sample, with respect to their
equilibrium positions, at $t=0$ \cite{Ashcroft}.

In Eq. (\ref{bPn}), $J_{P}(t)=\left\vert J_{P}(t)\right\vert \exp (i\nu
_{t}\pi )$ is the Jacobian factor given by Eq. (\ref{Jp}), $\varphi _{t}$ is
the SIVR phase at the time $t$ [Eq. (\ref{fitot})], and $\mathbf{Q}=\mathbf{K%
}_{f}-\mathbf{K}_{i}$ is the projectile momentum transfer. The function $\
\mathcal{V}_{n}^{^{(\mathbf{k},l)}}(\mathbf{R}_{t})$ is a crystal factor
that depends on the number $n$ of exchanged phonons. For zero-phonon
scattering, $\mathcal{V}_{0}^{^{(\mathbf{k},l)}}(\mathbf{R}_{t})$ is
independent on $(\mathbf{k},l)$ and can be expressed as
\begin{eqnarray}
\mathcal{V}_{0}(\mathbf{R}_{t}) &=&\int d\mathbf{q}\sum\limits_{\mathbf{r}_{%
\mathrm{B}}}\tilde{v}_{\mathbf{r}_{\mathrm{B}}}(\mathbf{q})\mathbf{\exp }%
\left[ -W_{\mathbf{r}_{\mathrm{B}}}\mathbf{(q)}\right]  \notag \\
&&\times \exp \left[ i\mathbf{q}\cdot \left( \mathbf{R}_{t}-\mathbf{r}_{%
\mathrm{B}}\right) \right] ,  \label{Vefn0}
\end{eqnarray}%
where $\tilde{v}_{\mathbf{r}_{\mathrm{B}}}(\mathbf{q})$ denotes the Fourier
transform of the binary interaction between the projectile and the crystal
ion placed at the Bravais position $\mathbf{r}_{\mathrm{B}}$, with $v_{%
\mathbf{r}_{\mathrm{B}}}$ coming from Eq. (\ref{Vps}). The summation on $%
\mathbf{r}_{\mathrm{B}}$ covers all the occupied lattice sites and $W_{%
\mathbf{r}_{\mathrm{B}}}\mathbf{(q)}$ is the usual momentum-dependent
Debye-Waller function, defined by Eq. (\ref{DWfactor}).

For one-phonon scattering, instead, $\mathcal{V}_{1}^{(\mathbf{k},l)}(%
\mathbf{R}_{t})$ depends on $(\mathbf{k},l)$, reading
\begin{eqnarray}
\mathcal{V}_{1}^{(\mathbf{k},l)}(\mathbf{R}_{t}) &=&\int d\mathbf{q\ }\left[
\mathbf{q}\cdot \mathbf{\alpha }_{l}(\mathbf{k})\right] \sum\limits_{\mathbf{%
r}_{\mathrm{B}}}\tilde{v}_{\mathbf{r}_{\mathrm{B}}}(\mathbf{q})\mathbf{\exp }%
\left[ -W_{\mathbf{r}_{\mathrm{B}}}\mathbf{(q)}\right]  \notag \\
&&\times \exp \left[ i\mathbf{q}\cdot \mathbf{R}_{t}\mathbf{+}i(\mathbf{k-q)}%
\cdot \mathbf{r}_{\mathrm{B}}\right] ,  \label{Vefn1}
\end{eqnarray}%
where $\mathbf{\alpha }_{l}(\mathbf{k})$ is the polarization vector
corresponding to the $(\mathbf{k},l)$ phonon.

From Eq. (\ref{Vefn0}) it can be noted that in absence of the Debye-Waller
factor, $\mathbf{\exp }\left[ -W_{\mathbf{r}_{\mathrm{B}}}\mathbf{(q)}\right]
$, $\mathcal{V}_{0}(\mathbf{R}_{t})$ coincides with the projectile-surface
potential for an ideal crystal, given by Eq. (\ref{Vps}) with $\underline{%
\mathbf{u}}=0$. Therefore, within the P0-SIVR approach the contribution of
the thermal lattice vibrations can be seen as an effective screening of the
projectile-surface interaction, given by the Debye-Waller factor, in
addition to the thermal effect on the projectile trajectories that is
produced by the different crystal configurations $\underline{\mathbf{u}}_{o}$%
.

\section{Results}

In this article we investigate the influence of the thermal lattice
vibrations on GIFAD patterns produced by $^{20}$Ne atoms grazingly colliding
with a LiF(001) surface at room temperature. Incidence along the $%
\left\langle 110\right\rangle $ direction, for which experimental spectra
were reported \cite{Gravielle2011}, is analyzed. Concerning the atomic
projectile, the relatively large mass of neon is expected to play some role
in inelastic processes, like phonon excitations \cite{Taleb2017,Roncin2018}.
However, we confine our study to the P0-SIVR approach, corresponding to
zero-phonon scattering, leaving the investigation of one-phonon excitations,
as given by Eq. (\ref{prob-P1-sivr}), for a future work.

\medskip

The P0-SIVR probability for scattering in the direction of the solid angle $%
\Omega _{f}=(\theta _{f},\varphi _{f})$ was derived from Eq. (\ref%
{prob-P0-sivr}) as
\begin{equation}
dP_{0}/d\Omega _{f}=K_{f}^{2}\left\vert \mathcal{A}_{n_{ph}=0}\right\vert
^{2},  \label{dP0}
\end{equation}%
where $\theta _{f}$ is the final polar angle, measured with respect to the
surface, and $\varphi _{f}$ is the azimuthal angle, measured with respect to
the axial channel. The transition amplitude $\mathcal{A}_{n_{ph}=0}$ was
calculated from Eq. (\ref{APn}), where the integration on $\mathbf{R}_{o}$
was reduced to the plane parallel to the surface, $\mathbf{R}_{o\Vert }$, by
considering that at $t=0$ all the classical trajectories start at a fixed
distance $Z_{o}$ from the surface, chosen as equal to the lattice constant,
for which the projectile is hardly affected by the surface interaction \cite%
{Gravielle2014, Gravielle2015}. In turn, the integral on \ $\mathbf{K}_{o}$
was solved in terms of the solid angle $\Omega _{o}=(\theta _{o},\varphi
_{o})$ that determines the $\mathbf{K}_{o}$- orientation, with $K_{o}=K_{i}$
accounting for the negligible energy dispersion of the incident beam \cite%
{Seifert2015,Gravielle2015}. In Eq. (\ref{APn}), the wave-packet profiles $f(%
\mathbf{R}_{o\Vert })$ and $g(\Omega _{o})$ were represented by products of
Gaussian functions, as respectively given by Eqs. (12) and (14) of Ref. \cite%
{Gravielle2015}. The widths of these profiles depend on the collimating
setup and the incidence conditions \cite{Gravielle2015,Gravielle2018}.
However, in this work we have used fixed values for such dispersion widths
in order to control their influence on the GIFAD patterns. Specifically, in
the Subsections III. A and III. B the angular widths were chosen as $\Delta
\theta _{o}=\Delta \varphi _{o}=0.03\deg $, values that are in the range of
the experimental conditions \cite{Gravielle2011}.

The projectile-surface interaction was described with the pairwise additive
model of Ref. \cite{Miraglia2017}. In addition, the integral on $\underline{%
\mathbf{u}}_{o}$ involved in Eq. (\ref{APn}) was evaluated by considering
that each crystal ion is randomly displaced from its equilibrium position
following an independent Gaussian distribution, with a mean-square
vibrational amplitude $\left\langle \mathbf{u}(\mathbf{r}_{\mathrm{B}%
})^{2}\right\rangle $. For the LiF(001) target at temperature $T=300$ K, the
$\left\langle \mathbf{u}(\mathbf{r}_{\mathrm{B}})^{2}\right\rangle $ values
were extracted from Ref. \cite{Schuller2010} by taking into account the
differences between Li and F ions and between bulk and surface (topmost
layer) sites.

For the calculation of $\mathcal{V}_{0}(\mathbf{R}_{t})$ [Eq. (\ref{Vefn0}%
)], the Debye-Waller function was approximated as $W_{\mathbf{r}_{\mathrm{B}%
}}\mathbf{(q)\simeq }q^{2}\left\langle \mathbf{u}(\mathbf{r}_{\mathrm{B}%
})^{2}\right\rangle /2$ \cite{Ashcroft,Roncin2017}. This assumption allowed
us to transform the $\mathbf{q}$- integral involved in Eq. (\ref{Vefn0})
into a space integral, which was solved together with the $\mathbf{R}%
_{o\Vert }$, $\Omega _{o}$ and $\underline{\mathbf{u}}_{o}$ integrals of Eq.
(\ref{APn}) by\ employing the MonteCarlo technique, with more than $6\times
10^{6}$ points for each incidence condition. Furthermore, for each
integration point the time integral involved in Eq. (\ref{bPn}) was
numerically solved by using an adaptive-stepsize method, with an error lower
than $1\%$. In this respect, the incorporation of $\underline{\mathbf{u}}%
_{o} $ into the evaluation of the projectile trajectories leads to increase
strongly the numerical effort necessary to reach the convergency of the
MonteCarlo integration, in relation to that required within the SIVR
approach \cite{Gravielle2014}.

\subsection{Thermal influence on GIFAD patterns}

\begin{figure}[tbp]
\includegraphics[width=0.5 \textwidth]{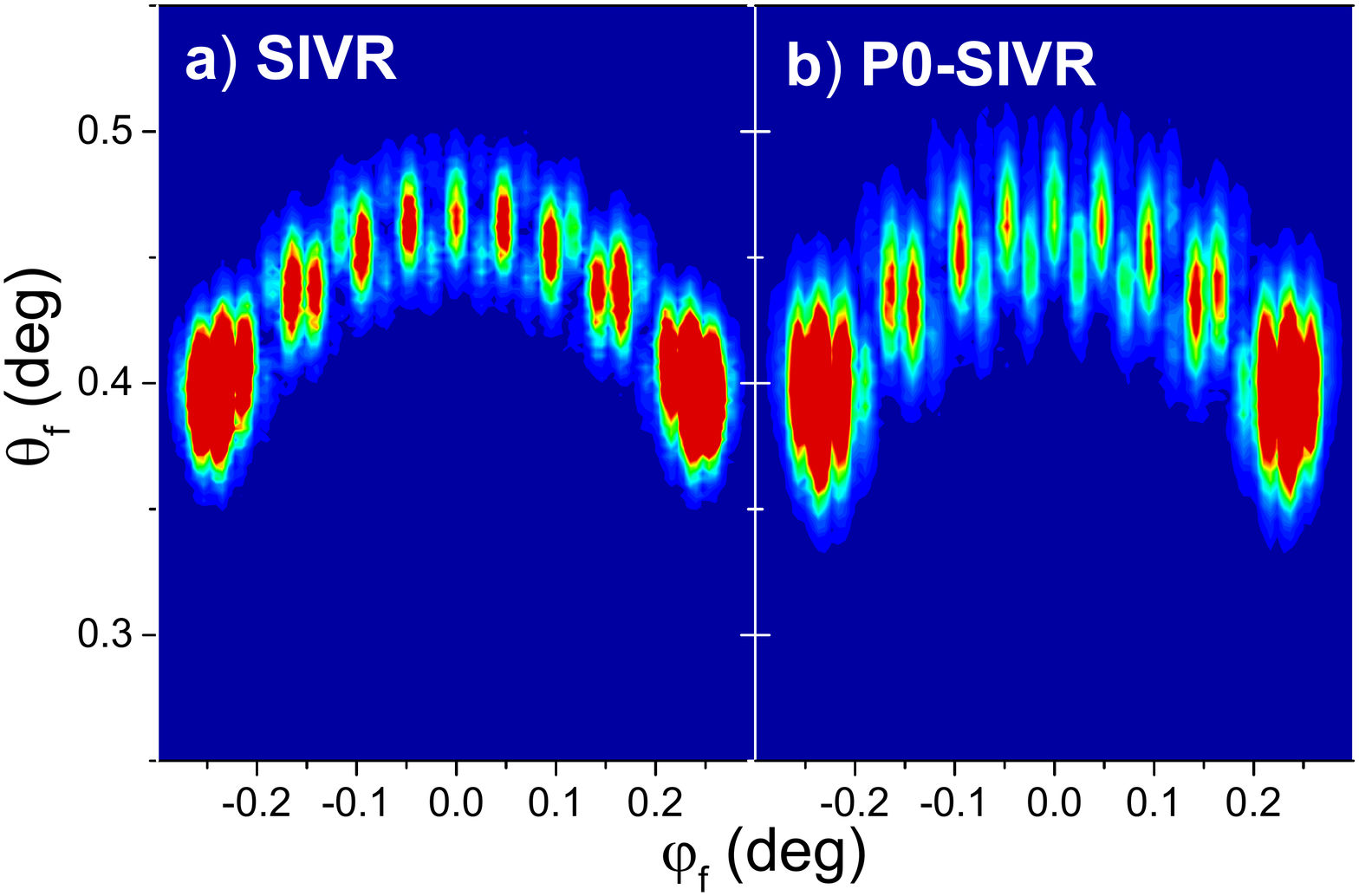} \centering
\caption{(Color online) Two-dimensional projectile distributions, as a
function of $\protect\theta _{f}$ and $\protect\varphi _{f}$, for Ne atoms
impinging on LiF(001) along the $\left\langle 110\right\rangle $ channel,
with $E=3.0$ keV and $\protect\theta _{i}=0.47\deg $. Results derived within
a) the SIVR approximation, for a static crystal, and b) the P$_{0}$-SIVR
approach, including thermal vibrations, are displayed.}
\label{map-e3ep02}
\end{figure}

In Fig. \ref{map-e3ep02} we show SIVR and P0-SIVR two-dimensional (2D)
distributions, as a function of the final angles $\theta _{f}$ and $\varphi
_{f}$, for Ne atoms impinging with the kinetic energy $%
E=K_{i}^{2}/(2m_{P})=3.0$ keV, $m_{P}$ being the projectile mass, and the
incidence angle $\theta _{i}=0.47\deg $, measured with respect to the
surface plane. \ \ Results for zero-phonon scattering derived within the
P0-SIVR approximation, displayed in the right panel of Fig. \ref{map-e3ep02}%
, include the phonon contribution, while the SIVR distribution, shown in the
left panel,\ was obtained by considering a static LiF crystal, with its ions
at rest at their equilibrium positions \cite{Gravielle2014}. Within both
approaches, the width of the spatial profile $f(\mathbf{R}_{o\Vert })$ was
chosen to cover two equivalent parallel channels, which\ gives rise to Bragg
maxima produced by inter- channel interference \cite{Gravielle2015}.

In Figs. \ref{map-e3ep02} a) and \ref{map-e3ep02} b) the Bragg maxima look
like vertical strips placed inside an annulus with mean radius $\theta _{i}$%
, due to the energy conservation. Even though the SIVR and P0-SIVR
distributions of Fig. \ref{map-e3ep02} display qualitatively similar
interference patterns, with almost the same $\varphi _{f}$- extension of the
spectrum, the relative intensities of the interference maxima, as well as
the $\theta _{f}$- angular spreads, predicted by the two approximations
differ each other, these discrepancies being indicative of the effect of \
the thermal lattice vibrations.

\begin{figure}[tbp]
\includegraphics[width=0.5\textwidth]{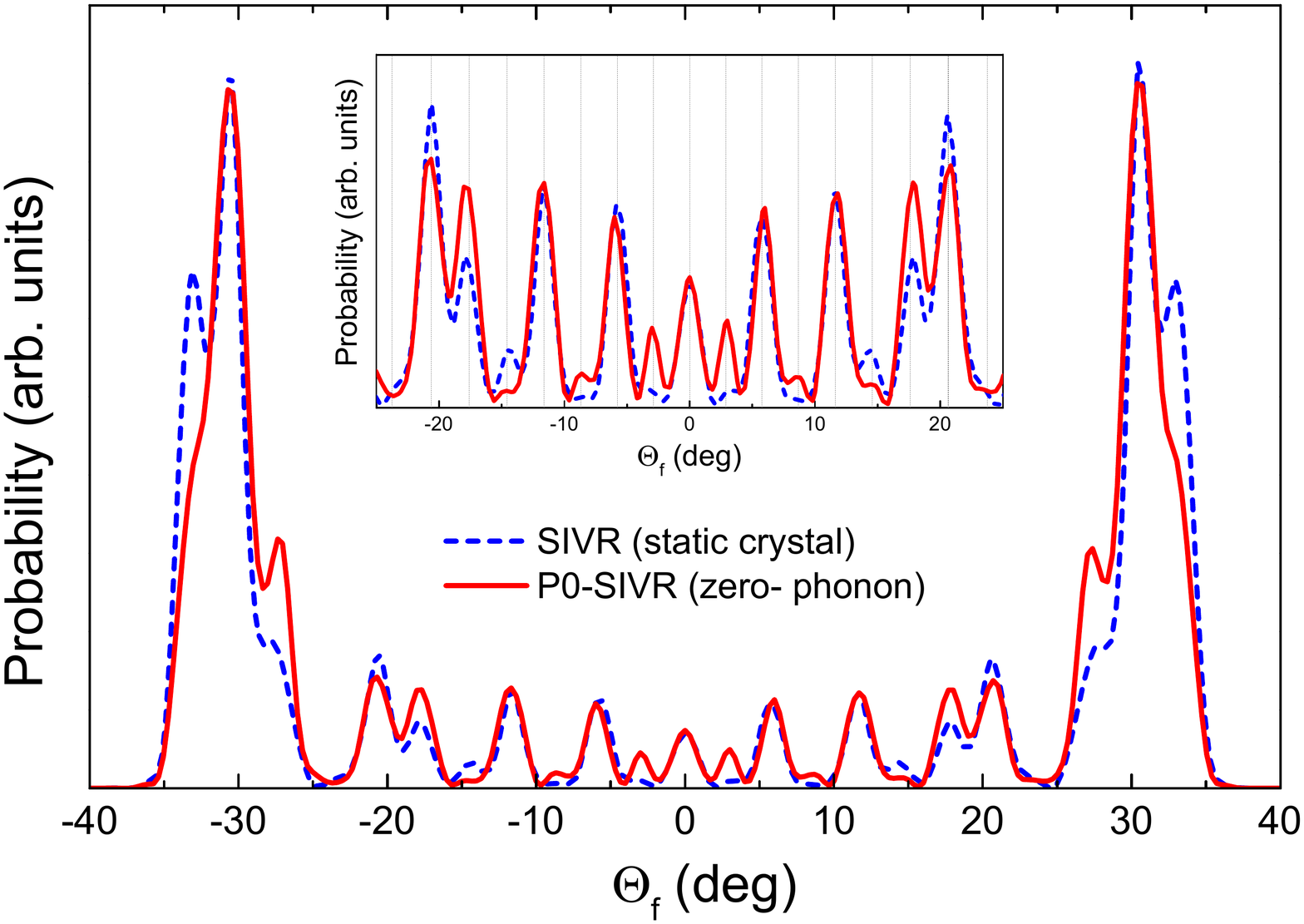} \centering
\caption{(Color online) Differential probability, as a function of the
deflection angle $\Theta _{f}$, for the case of Fig. \protect\ref{map-e3ep02}%
. Red solid line, zero-phonon scattering probability derived within the P$%
_{0}$-SIVR approach; blue dashed line, SIVR probability for a static
crystal. The inset displays a zoomed view of the central region of the
spectrum. Dashed vertical lines indicate Bragg-peak positions.}
\label{bragg-ep02}
\end{figure}

To look with more detail into the projectile distributions of Fig. \ref%
{map-e3ep02}, in Fig. \ref{bragg-ep02} we plot the corresponding $%
dP_{0}/d\Theta _{f}$ probabilities, as a function of the deflection angle $%
\Theta _{f}=\arctan (\varphi _{f}/\theta _{f})$. These differential
probabilities were obtained by integrating Eq. (\ref{dP0}) over a reduced
annulus of mean radius $\theta _{i}$ and central thickness $0.03\deg $, as
it is usually done to derive the experimental projected intensities \cite%
{Winter2011,Debiossac2016}. From Fig. \ref{bragg-ep02} we observe that the
angular positions of the Bragg peaks (indicated with vertical lines in the
inset) are not affected by the thermal vibrations, coinciding for the
P0-SIVR and SIVR approximations. Instead, the relative intensities of the
Bragg maxima are strongly modified by the contribution of the thermal
fluctuations included in the P0-SIVR approach, which can increase or reduce
the SIVR intensity of a given Bragg order, as shown in the inset of Fig. \ref%
{bragg-ep02}. Hence, since the use of GIFAD \ for surface analysis is
commonly based on the comparison of the relative intensities of the
interference maxima with theoretical models, these results suggest that the
thermal vibrations might play an important role in the GIFAD\ technique.

At this point, it is important to take into account that the Bragg-peak
intensities are determined by an intra-channel factor due to the
interference inside a single channel, which acts as an enveloped function of
the inter-channel interference \cite{Gravielle2014}. Therefore, for the
purpose of analyzing the influence of lattice vibrations on the Bragg
intensities under different incidence conditions, hereinafter we restrict
our study to pure intra-channel spectra, which are produced by initial
wave-packet \ profiles covering a transverse distance equal to the channel
width \cite{Gravielle2018}.

\subsection{Thermal effects in the intra-channel interference}

\begin{figure}[tbp]
\includegraphics[width=0.5\textwidth]{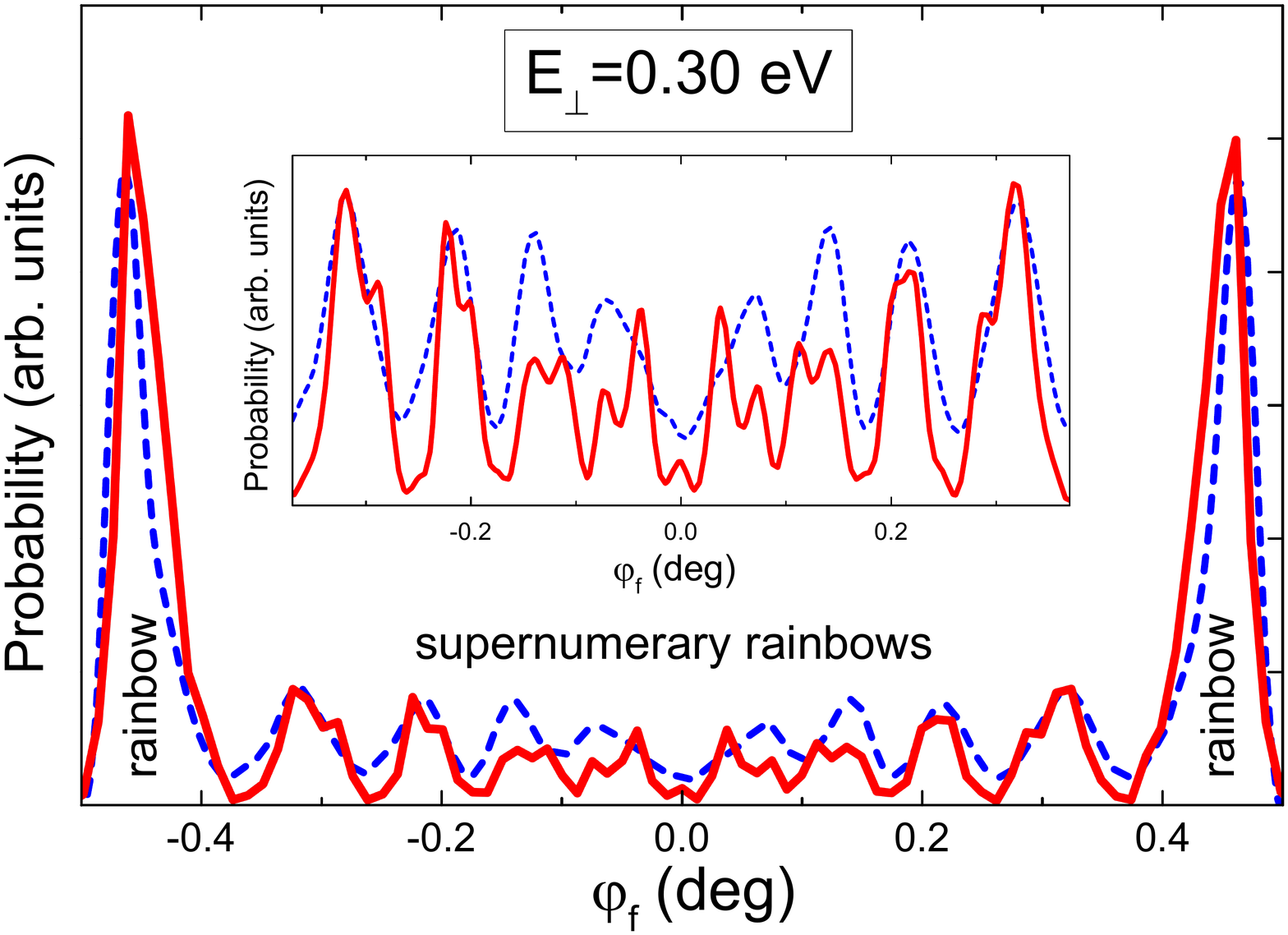} \centering
\caption{(Color online) Intra-channel distribution, as a function of the
final azimuthal angle $\protect\varphi _{f}$, for $E=1.3$ keV and the normal
energy $E_{\bot }=0.30$ eV. Lines, analogous to Fig. \protect\ref{bragg-ep02}%
. Inset: Zoomed view of the central region.}
\label{spectra-ep03}
\end{figure}

GIFAD distributions due to a single coherently illuminated channel are
governed by the normal incidence energy, $E_{\perp }=E\sin ^{2}\theta _{i}$,
which is associated with the projectile motion perpendicular to the surface
plane \cite{Schuller2009,Winter2011}. In Fig. \ref{spectra-ep03} we display
P0-SIVR and SIVR intra-channel spectra, as a function of the azimuthal angle
$\varphi _{f}$, for $E=1.3$ keV and $E_{\perp }=0.30$ eV. Notice that this
normal energy is barely lower than the upper $E_{\perp }$- limit of
available GIFAD experiments for Ne/LiF(001) \cite{Gravielle2011}. Both
curves of Fig. \ref{spectra-ep03} display equivalent interference patterns,
with rainbow and supernumerary rainbow maxima. While the rainbow peaks,
corresponding to the high-intensity outermost maxima, have a classical
origin, the supernumerary peaks are produced by quantum interference, being
expected to be more affected by thermal fluctuations than the rainbow, which
is confirmed in Fig. \ref{spectra-ep03}.

The P0-SIVR spectrum of Fig. \ref{spectra-ep03} presents a small angular
shift in the positions of the central supernumeraries, with respect to those
corresponding to the SIVR curve. But in addition, it is observed that the
thermal vibrations included in the P0-SIVR approach affect the shape and
relative intensity of the supernumerary peaks, specially in the central
region of the spectrum, where a noticeable double-peak structure is clearly
visible in each supernumerary maxima (see the inset of Fig. \ref%
{spectra-ep03}). \ Remarkably, this interference sub-pattern that appears as
a superimposed structure on the P0-SIVR supernumeraries is mainly produced
by the effect of the thermal deviations $\underline{\mathbf{u}}_{o}$ on the
projectile trajectories. When the $\mathbf{R}_{t}$- dependence on $%
\underline{\mathbf{u}}_{o}$ is left aside, results derived from Eq. (\ref%
{prob-P0-sivr}) by considering an ideal and static crystal, but keeping the
factor $\exp \left[ -W_{\mathbf{r}_{\mathrm{B}}}\mathbf{(q)}\right] $ in Eq.
(\ref{Vefn0}), fully agree with the SIVR values, indicating that the
Debye-Waller factor plays a minor role in the elastic scattering at room
temperature.

\begin{figure}[tbp]
\includegraphics[width=0.5 \textwidth]{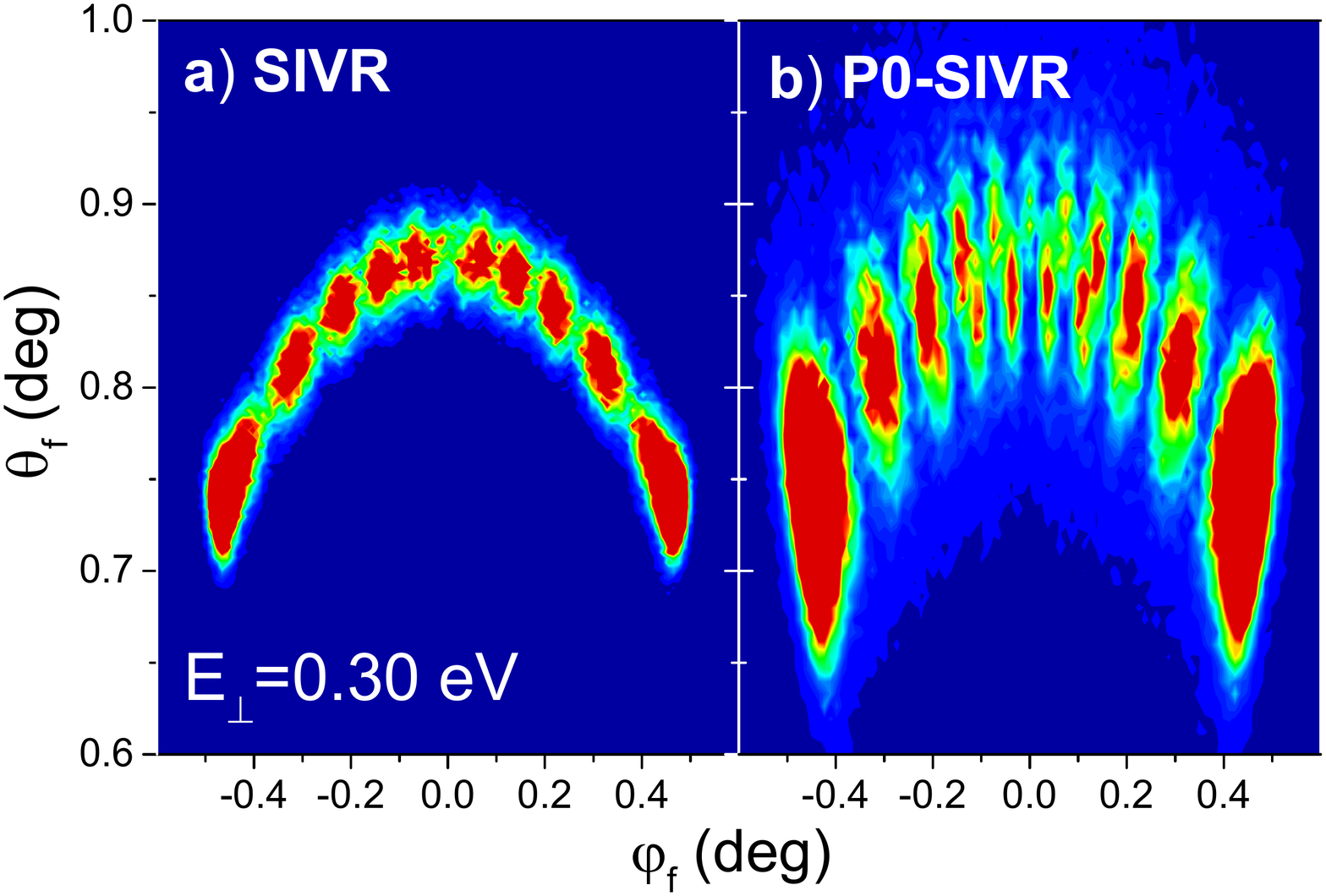} \centering
\caption{(Color online) Analogous to Fig. \protect\ref{map-e3ep02} for the
case of Fig. \protect\ref{spectra-ep03}, i.e. $E=1.3$ keV and $E_{\bot
}=0.30 $ eV. }
\label{map-ep03}
\end{figure}

To understand the origin of the interference sub-patterns observed \ in Fig. %
\ref{spectra-ep03}, the corresponding 2D- angular distributions, as a
function of $\theta _{f}$ and $\varphi _{f}$, are plotted in Fig. \ref%
{map-ep03}. In Fig. \ref{map-ep03} a) the SIVR distribution for the static
crystal displays broad interference maxima, which lay on an annulus whose
thickness is essentially determined by the polar-angle dispersion $\Delta
\theta _{o}$ of the atomic beam \cite{Gravielle2015,Gravielle2016}. Instead,
in Fig. \ref{map-ep03} b) the thermal lattice vibrations introduce an
additional polar-angle spread into the P0-SIVR distribution, transforming
the SIVR interference spots into vertical strips. The emergence of a
polar-angle broadening as a consequence of thermal fluctuations was already
proposed in Ref. \cite{Manson2008}. Furthermore, it is found that the
thermal vibrations give rise to an interference structure in the P0-SIVR
distribution along $\theta _{f}$, which is more evident around $\varphi
_{f}\approx 0$. Then, the double-peak shape of the internal P0-SIVR maxima
of Fig. \ref{spectra-ep03} corresponds to the projected image on $\varphi
_{f}$ of such a vertical pattern \cite{Manson2008}, which is produced by the
interference among projectiles running nearly on top of thermally-shifted Li
and F rows \cite{Schuller2010}.

\subsection{Experimental comparison}

In order to test the reliability of the proposed model, in Fig. \ref%
{spectrum-expto} we contrast P0-SIVR and SIVR differential probabilities, as
a function of the deflection angle $\Theta _{f}$, with the available
experimental spectrum \cite{Gravielle2011} for the incidence conditions $%
E=1.3$ keV and $\theta _{i}=0.55\deg $, which correspond to the normal
energy $E_{\perp }=0.12$ eV. Like in the previous Subsection, in this case
the theoretical and experimental distributions display only supernumerary
peaks, associated with intra-channel interference, without any trace of
Bragg interference.

\begin{figure}[tbp]
\includegraphics[width=0.5\textwidth]{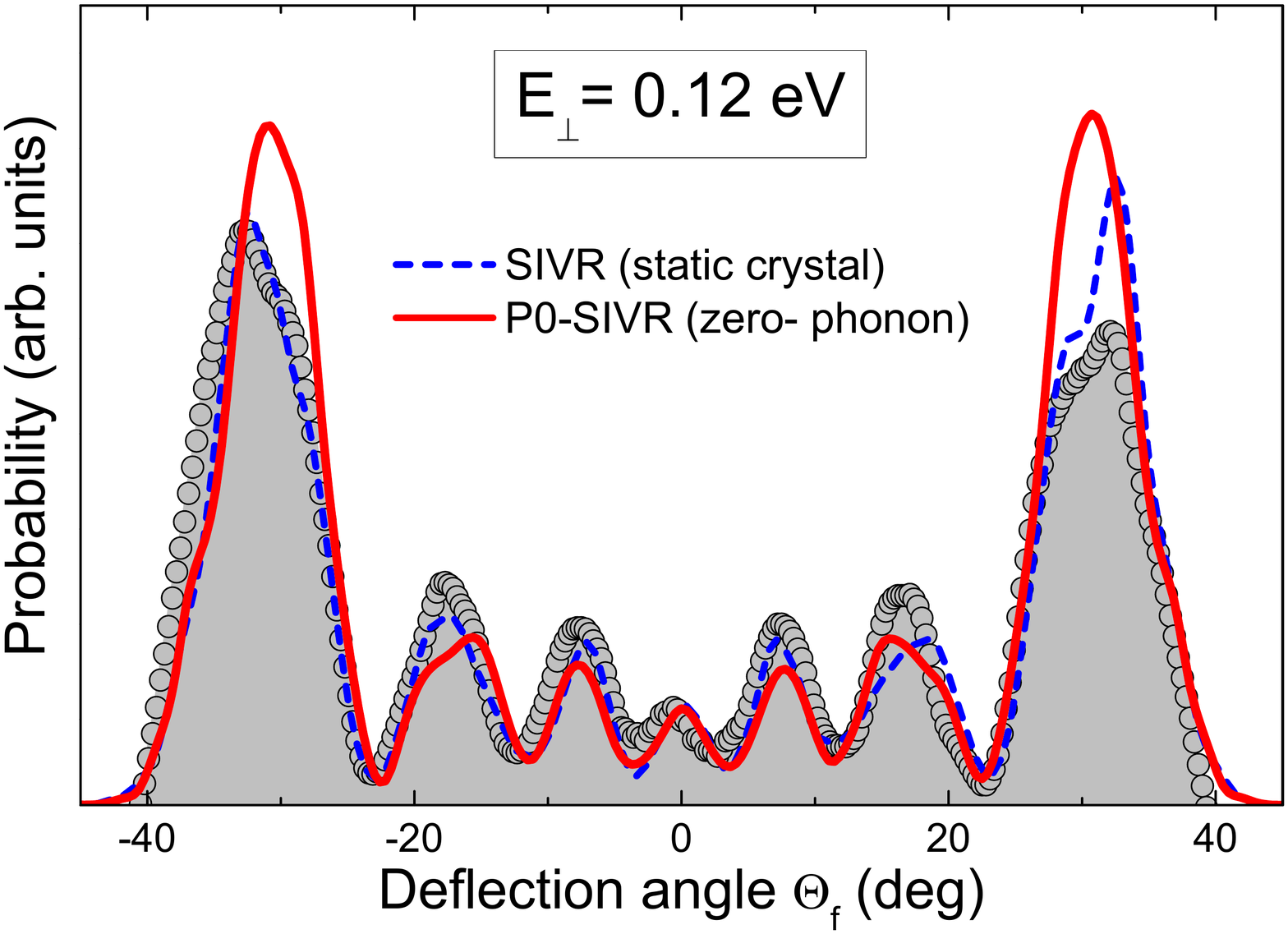} \centering
\caption{(Color online)Angular distribution, as a function of the deflection
angle $\Theta _{f}$, for the incidence energy $E=1.3$ keV and angle $\protect%
\theta _{i}=0.55\deg $ [i.e., $E_{\perp }=0.12$ eV]. Lines, analogous to
Fig. \protect\ref{bragg-ep02}; gray solid circles, experimental data from
Ref. \protect\cite{Gravielle2011}.}
\label{spectrum-expto}
\end{figure}

In Fig. \ref{spectrum-expto}, the P0-SIVR spectrum is very similar to that
for a static crystal derived by means of the SIVR approach, both showing a
very good agreement with the experimental data. This behavior, together with
the absence of interference sub-structures in the P0-SIVR supernumeraries,
might indicate that the thermal contribution on GIFAD patterns \ becomes
smaller as $E_{\perp }$ decreases, since the atomic projectiles move farther
from the surface plane.

\begin{figure}[tbp]
\includegraphics[width=0.5 \textwidth]{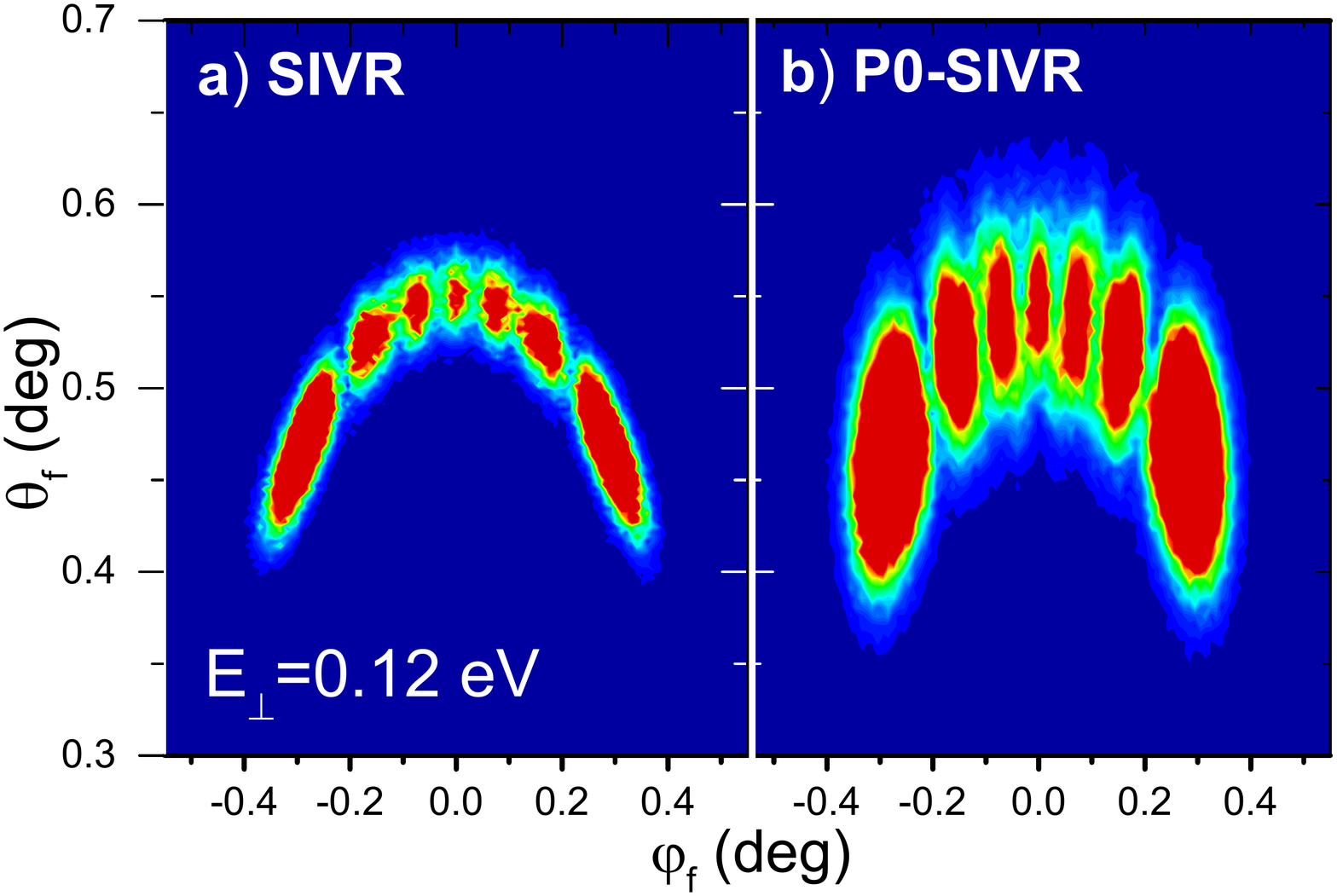} \centering
\caption{(Color online) Analogous to Fig. \protect\ref{map-e3ep02} for the
case of Fig. \protect\ref{spectrum-expto}, i.e. $E=1.3$ keV and $E_{\bot
}=0.12$ eV. }
\label{map-expto}
\end{figure}

However, notice that the $\Theta _{f}$- spectra displayed in Fig. \ref%
{spectrum-expto} were also obtained by integrating the corresponding 2D-
angular distributions, shown in Fig. \ref{map-expto}, inside an annulus of
central thickness $0.03\deg $ \cite{Gravielle2011,Debiossac2016}. When the
distributions of Figs. \ref{map-expto} a) and \ref{map-expto} b) are
compared, it is found that even though for this low perpendicular energy
there are no visible signatures of interference sub-structures, the thermal
motion of the crystal ions still introduces a wide polar-angle dispersion in
the P0-SIVR distribution. It gives rise to a P0-SIVR pattern formed by
elongated vertical streaks, instead of the nearly circular spots of the SIVR
distribution, which is in good accord with the experiment of Fig. 1 a) of
Ref. \cite{Gravielle2011}.

\subsection{Incoherent thermal vibrations}

\begin{figure}[tbp]
\includegraphics[width=0.5 \textwidth]{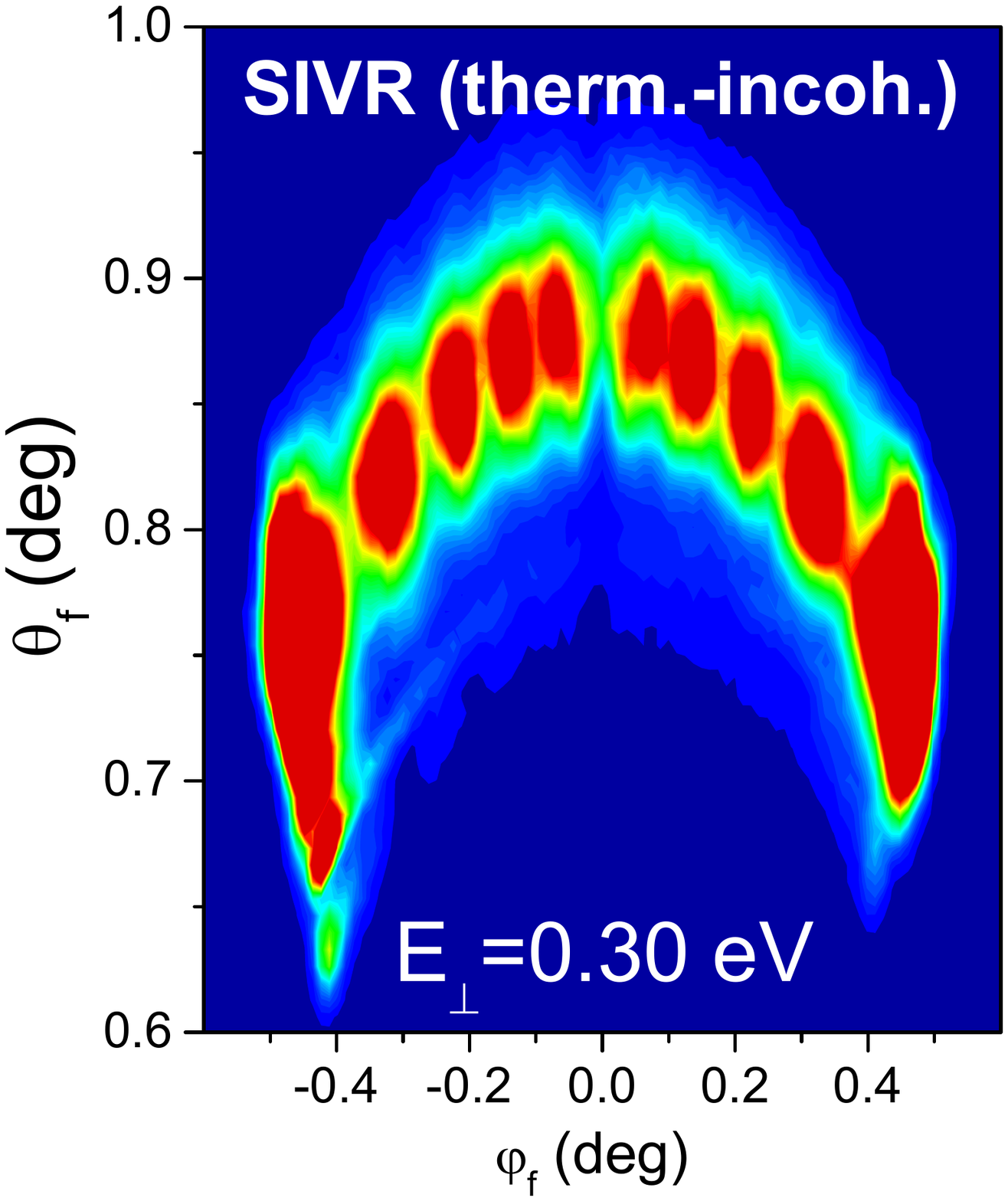}
\centering \caption{(Color online) Two-dimensional projectile
distribution, as a function of $\protect\theta _{f}$ and
$\protect\varphi _{f}$, evaluated with the thermally-incoherent SIVR
model, as explained in the text. Results for the case of Fig.
\protect\ref{map-ep03}, corresponding to $E_{\bot }=0.30$ eV, are
displayed. } \label{incohvib}
\end{figure}

Lastly, we investigate the role of the coherent thermal contribution
involved in the present approach by comparing the P0-SIVR results with
thermally-incoherent probabilities derived from the SIVR approach. Such an
incoherent calculation was done by averaging the SIVR probability, i.e., the
square modulus of the SIVR transition amplitude \cite{Gravielle2014}, for
different configurations $\underline{\mathbf{u}}_{o}$ of the crystal target,
where the crystal ions are randomly displaced from their equilibrium
positions following Gaussian distributions, as considered within the P0-SIVR
model. In Fig. \ref{incohvib} we plot the thermally-incoherent SIVR
distribution, as a function of $\theta _{f}$ and $\varphi _{f}$, for the
case of Fig. \ref{map-ep03} corresponding to the normal energy $E_{\perp
}=0.30$ eV. By contrasting Figs. \ref{map-ep03} b) and \ref{incohvib} we
found that the incoherent addition of thermal effects destroys the
interference sub-pattern observed in the central region of P0-SIVR
distribution of Fig. \ref{map-ep03} b). Moreover, the thermally-incoherent
approach introduces a significant broadening of the supernumerary maxima
along $\varphi _{f}$, while the $\theta _{f}$- dispersion is similar to that
displayed by the P0-SIVR distribution. Although there is no available
experimental distribution for this case, an analogous calculation for the
case of Fig. \ref{map-expto} shows that this noticeable $\varphi _{f}$-
widening of the interference peaks, associated with the incoherent thermal
contribution, does not agree with the reported experiments \cite%
{Gravielle2011}. Therefore, the thermally-incoherent SIVR approximation \cite%
{Miraglia2017} seems not to be suitable to reproduce thermal effects on
GIFAD patterns.

\section{Conclusions}

We have developed the P-SIVR approximation, which is a semi-quantum method
that takes into account the contribution of the vibrational modes of the
crystal to the GIFAD patterns. The P-SIVR probability was expressed as a sum
of partial scattering probabilities, P$n$-SIVR, each of them involving the
exchange of a different number $n$ of phonons. Formulas for the
probabilities corresponding to zero- and one- phonon scattering have been
presented.

The P0-SIVR approach for zero-phonon scattering was employed to investigate
the effect of the thermal lattice vibrations on GIFAD distributions for the
Ne/LiF(001) system. At room temperature it was found that, depending on the
incidence conditions, the relative intensities of the Bragg peaks can be
affected by the thermal fluctuations of the LiF(001) crystal. Within the
P0-SIVR model, the thermal vibrations introduce a polar-angle dispersion
into the angular distributions, which contribute to transform the
interference spots into elongated structures, in concordance with the
experimental observations \cite{Gravielle2011}. \ \ For high normal
energies, such a polar-angle spread can also alter the shape of the
supernumerary maxima, giving rise to the appearance of interference sub-
patterns in the central region of the GIFAD spectra.

In conclusion, present P0-SIVR results demonstrate that thermal vibrations
affect the aspect of the GIFAD patterns from insulator surfaces, a finding
that is especially relevant for the use of GIFAD as a surface analysis
technique. But notice that there are other effects, like phonon excitations
\cite{Roncin2017,Roncin2018} or the presence of terraces in the crystal
sample \cite{Lalmi2012}, not considered in this article, which can modify
the interference structures too. Therefore, further experimental and
theoretical work to investigate the different decoherence mechanisms in
GIFAD would be valuable.

\begin{acknowledgments}
The authors acknowledge financial support from CONICET and ANPCyT of
Argentina.
\end{acknowledgments}

\appendix*

\section{P-SIVR model for grazing atom-surface scattering with phonon
exchange}

In this Appendix we explain the steps and assumptions that lead to the
P-SIVR approximation for GIFAD from an insulator target. Let us consider an
atomic projectile ($P$), with initial momentum $\mathbf{K}_{i}$, which is
scattered from a crystal surface ($S$), ending in a final state with
momentum $\mathbf{K}_{f}$. The scattering state of the projectile-surface
system at the time $t$, $\left\vert \Psi _{i}(t)\right\rangle $, is governed
by the time-dependent Schr\"{o}dinger equation
\begin{equation}
\left[ \frac{\mathbf{P}_{P}^{2}}{2m_{P}}+H_{S}+V_{PS}\right] \left\vert \Psi
_{i}(t)\right\rangle =i\frac{d}{dt}\left\vert \Psi _{i}(t)\right\rangle ,
\label{H}
\end{equation}%
where $\mathbf{P}_{P}\ $denotes the momentum operator of the projectile with
mass $m_{P}$, $H_{S}$ is the unperturbed surface Hamiltonian, and $V_{PS}$
is the perturbation produced by the projectile-surface interaction. The
Hamiltonian $H_{S}$ reads
\begin{equation}
H_{S}=\sum_{\mathbf{r}_{\mathrm{B}}}\frac{\mathbf{P}^{2}(\mathbf{r}_{\mathrm{%
B}})}{2m(\mathbf{r}_{\mathrm{B}})}+\mathcal{W}_{S}(\underline{\mathbf{u}}),
\label{Hs}
\end{equation}%
where the sum runs over the positions $\mathbf{r}_{\mathrm{B}}$ of the
occupied Bravais lattice sites. In Eq. (\ref{Hs}) $\mathbf{P}(\mathbf{r}_{%
\mathrm{B}})$ indicates the momentum operator of the crystal ion that
oscillates about $\mathbf{r}_{\mathrm{B}}$ and $m(\mathbf{r}_{\mathrm{B}})$
is its mass, with $m(\mathbf{r}_{\mathrm{B}})=$ $m_{1}$ or $m_{2}$ to
include two different ions in the crystallographic basis. The potential $%
\mathcal{W}_{S}(\underline{\mathbf{u}})$ represents the potential energy of
the crystal as a function of the multi-dimensional vector $\underline{%
\mathbf{u}}$, which is determined by the spatial deviations $\mathbf{u}(%
\mathbf{r}_{\mathrm{B}})$ of the crystal ions from their equilibrium
positions $\mathbf{r}_{\mathrm{B}}$, for all the occupied lattice sites \cite%
{Ashcroft}.

\bigskip

As initial condition, at $t=0$, when the projectile is far away from the
surface, the scattering state $\left\vert \Psi _{i}(t)\right\rangle $ tends
to the state $\left\vert \chi _{i}(0)\right\rangle $, where
\begin{equation}
\chi _{j}(t)=e^{i\mathbf{K}_{j}\cdot \mathbf{R}_{P}}\phi _{j}(\underline{%
\mathbf{u}})\exp (-iE_{j}t),\quad j=i(f),  \label{fi-j}
\end{equation}%
is the initial (final) unperturbed wave function with total energy
\begin{equation}
E_{j}=K_{j}^{2}/(2m_{P})+\epsilon _{j},\quad j=i(f),  \label{E}
\end{equation}%
which satisfies the energy conservation, i.e., $E_{i}=E_{f}$. In Eq. (\ref%
{fi-j}), $\mathbf{R}_{P}$ is the position vector of the center of mass of
the incident atom and the wave function $\phi _{j}(\underline{\mathbf{u}})$,
for $j=i(f)$, is the initial (final) eigenstate of $H_{S}$ with eigenvalue $%
\epsilon _{j}$.

By considering that the surface behaves like a harmonic crystal, $H_{S}$ can
be expressed as a sum of independent harmonic-oscillator Hamiltonians, each
of them corresponding to a different normal mode of the lattice, with wave
vector $\mathbf{k\ }$, frequency $\omega _{l}(\mathbf{k})$, and $l$ denoting
the phonon branch. Hence, the unperturbed crystal state $\phi _{j}$, for $%
j=i,f$, is determined by the excitation numbers $n_{\mathbf{k},l}^{(j)}$ of
the normal modes and the corresponding crystal energy reads

\begin{equation}
\epsilon _{j}=\sum\limits_{\mathbf{k},l}\omega _{l}(\mathbf{k})\left[ n_{%
\mathbf{k},l}^{(j)}+\frac{1}{2}\right] ,\quad j=i,f,  \label{Hs-harm}
\end{equation}%
where the sum runs over all the $(\mathbf{k},l)$ normal modes of the crystal
\cite{Ashcroft}.

\subsection{P-SIVR scattering state}

Within the P-SIVR method, the scattering state $\left\vert \Psi
_{i}(t)\right\rangle $\ is approximated by means of the IVR method \cite%
{Miller2001}. It is expressed as

\begin{eqnarray}
\left\vert \Psi _{i}^{(\mathrm{P-SIVR})}(t)\right\rangle &=&\int d\mathbf{R}%
_{o}\ f(\mathbf{R}_{o})\int d\mathbf{K}_{o}\ g(\mathbf{K}_{o})  \notag \\
&&\times \int d\underline{\mathbf{u}}_{o}\int d\underline{\mathbf{p}}_{o}%
\left[ J(t)\right] ^{1/2}\exp (i\mathbf{K}_{i}\cdot \mathbf{R}_{o})  \notag
\\
&&\times \exp (iS_{t})\phi _{i}(\underline{\mathbf{u}}_{o})\left\vert
\mathbf{R}_{t}\right\rangle \otimes \left\vert \underline{\mathbf{u}}%
_{t}\right\rangle ,  \label{state-psivr}
\end{eqnarray}%
where the position ket $\left\vert \mathbf{R}_{t}\right\rangle $ is
associated with the time-evolved position $\mathbf{R}_{t}$ of the incident
atom at a given time $t$, which is derived by considering a classical
trajectory with starting position and momentum $\mathbf{R}_{o}$ and $\mathbf{%
K}_{o}$, respectively. In a similar way, the deviation ket $\left\vert
\underline{\mathbf{u}}_{t}\right\rangle $ is determined from the classical
deviations $\mathbf{u}_{t}(\mathbf{r}_{\mathrm{B}})$ of all the crystal
ions, starting at $t=0$ from initial deviations and momenta $\mathbf{u}_{o}(%
\mathbf{r}_{\mathrm{B}})$ and $\mathbf{p}_{o}(\mathbf{r}_{\mathrm{B}})$,
respectively. In Eq. (\ref{state-psivr}), $\underline{\mathbf{u}}_{o}$ ($%
\underline{\mathbf{p}}_{o}$) denotes the $3N$-dimension vector determined by
such deviations (momenta) for the $N$ ions contained in the crystal target.
In fact, note that we are dealing with a many-particle problem, in which the
classical motions of the projectile and the crystal ions are related through
their mutual interactions. Consequently, the classical trajectories $\mathbf{%
R}_{t}$ and $\mathbf{u}_{t}(\mathbf{r}_{\mathrm{B}})$, for the different $%
\mathbf{r}_{\mathrm{B}}$-values, depend on the initial positions and momenta
of all the particles in the system.

Furthermore, in Eq. (\ref{state-psivr}) the functions $f(\mathbf{R}_{o})$
and $g(\mathbf{K}_{o})$ describe the shape of the position- and momentum-
wave packet associated with the incident projectile, while $S_{t}$
represents the classical action along the trajectory, reading
\begin{eqnarray}
S_{t} &=&\int\limits_{0}^{t}dt^{\prime }\ \left[ \frac{\mathbf{K}_{t^{\prime
}}^{2}}{2m_{P}}-V_{PS}(\mathbf{R}_{t^{\prime }},\underline{\mathbf{u}}%
_{t^{\prime }})\right.  \notag \\
&&\left. +\sum\limits_{\mathbf{r}_{\mathrm{B}}}\frac{\mathbf{p}_{t^{\prime
}}^{2}(\mathbf{r}_{\mathrm{B}})}{2m(\mathbf{r}_{\mathrm{B}})}-\mathcal{W}%
_{S}\left( \underline{\mathbf{u}}_{t^{\prime }}\right) \right] ,  \label{St}
\end{eqnarray}%
where $\mathbf{K}_{t}=m_{P}d\mathbf{R}_{t}/dt\ $\ and $\mathbf{p}_{t}(%
\mathbf{r}_{\mathrm{B}})=m(\mathbf{r}_{\mathrm{B}})d\mathbf{u}_{t}(\mathbf{r}%
_{\mathrm{B}})/dt$ are the classical projectile and crystal ion momenta,
respectively, at the time $t$. The Jacobian factor
\begin{equation}
J(t)=\det \left[ \frac{\partial \mathbf{R}_{t}\partial \underline{\mathbf{u}}%
_{t}}{\partial \mathbf{K}_{o}\partial \underline{\mathbf{p}}_{o}}\right]
\label{J}
\end{equation}%
is a determinant evaluated along the classical path, which takes into
account the motions of the projectile and all the crystal ions. This
Jacobian factor can be related to the Maslov index \cite{Guantes2004} by
expressing it as $J(t)=\left\vert J(t)\right\vert \exp (i\nu _{t}\pi )$,
where $\left\vert J(t)\right\vert $ is the modulus of $J(t)$ and $\nu _{t}$
is an integer number that increases by $1$ every time that $J(t)$ changes
its sign along the time.

\subsection{P-SIVR transition amplitude}

By using the P-SIVR scattering state, given by Eq. \ (\ref{state-psivr}),
within the framework of the time-dependent distorted-wave formalism \cite%
{Dewangan1994}, the P-SIVR transition amplitude reads
\begin{equation}
A^{(\mathrm{P-SIVR})}=-i\int\limits_{0}^{+\infty }dt\left\langle \chi
_{f}\left( t\right) \left\vert V_{PS}\right\vert \Psi _{i}^{(\mathrm{P-SIVR}%
)}(t)\right\rangle .  \label{Adw}
\end{equation}%
For the evaluation of Eq. (\ref{Adw}) a meaningful simplification can be
obtained by considering that in GIFAD the interaction time of the projectile
with the crystal surface is much shorter than the characteristic time of
phonon vibrations \cite{Ashcroft}. Therefore, we can assume that the crystal
ions remain at their initial positions $\mathbf{u}_{o}(\mathbf{r}_{\mathrm{B}%
})$ during the collision, leading to

\begin{equation}
J(t)\approx J_{P}(t)=\det \left[ \frac{\partial \mathbf{R}_{t}}{\partial
\mathbf{K}_{o}}\right] .  \label{Jp}
\end{equation}%
Then, by introducing the closure relation for the initial deviations of the
crystal ions, the P-SIVR transition amplitude can be expressed, except for a
normalization factor,  as%
\begin{eqnarray}
A^{(\mathrm{P-SIVR})} &\equiv &A\left[ a_{if}\right] =\int d\mathbf{R}_{o}\
f(\mathbf{R}_{o})\int d\mathbf{K}_{o}\ g(\mathbf{K}_{o})  \notag \\
&&\times \ \int d\underline{\mathbf{u}}_{o}\ a_{if},  \label{Apsivr}
\end{eqnarray}%
where
\begin{eqnarray}
a_{if} &=&\int\limits_{0}^{+\infty }dt\ \left\vert J_{P}(t)\right\vert
^{1/2}e^{i\nu _{t}\pi /2}\ F_{if}^{(c)}(\mathbf{R}_{t},t)  \notag \\
&&\times \exp \left[ i\left( \varphi _{t}-\mathbf{Q}\cdot \mathbf{R}%
_{o}\right) \right] \quad   \label{aif}
\end{eqnarray}%
is the partial amplitude associated with the classical path $\mathbf{R}%
_{t}\equiv \mathbf{R}_{t}(\mathbf{R}_{o},\mathbf{K}_{o},\underline{\mathbf{u}%
}_{o})$, which was derived by assuming that the initial deviations $%
\underline{\mathbf{u}}_{o}$ are decoupled from $\phi _{i}(\underline{\mathbf{%
u}})$. \ In Eq. (\ref{aif}), the function $F_{if}^{(c)}$ is defined as
\begin{equation}
F_{if}^{(c)}(\mathbf{R}_{t},t)=\left\langle \Phi _{f}(t)\left\vert V_{PS}(%
\mathbf{R}_{t},\underline{\mathbf{u}})\right\vert \Phi _{i}(t)\right\rangle ,
\label{Vsp}
\end{equation}%
where $\Phi _{j}(t)=\phi _{j}(\underline{\mathbf{u}})\exp (-i\epsilon _{j}t)$%
, for $j=i,f$ , $\mathbf{Q}=\mathbf{K}_{f}-\mathbf{K}_{i}$, and
\begin{equation}
\varphi _{t}=\int\limits_{0}^{t}dt^{\prime }\ \left[ \frac{\left( \mathbf{K}%
_{f}-\mathbf{K}_{t^{\prime }}\right) ^{2}}{2m_{P}}-V_{PS}(\mathbf{R}%
_{t^{\prime }},\underline{\mathbf{u}}_{o})\right]   \label{fitot}
\end{equation}%
is the SIVR phase at the time $t$ \cite{Gravielle2014}. By contrasting Eq. (%
\ref{aif}) with the SIVR partial amplitude for a static surface, given by
Eq. (6) from Ref. \cite{Gravielle2015}, notice that, apart from the
dependence of $\mathbf{R}_{t}$ and $\varphi _{t}$ on $\underline{\mathbf{u}}%
_{o}$, the P0-SIVR partial amplitude differs from the SIVR one by the
substitution of the projectile-surface potential by the crystal factor $%
F_{if}^{(c)}$, which is related to the first-order Born amplitude for the
crystal-state transition $\left\vert \phi _{i}\right\rangle \rightarrow
\left\vert \phi _{f}\right\rangle $.

\subsection{P-SIVR differential probability}

The P-SIVR differential probability for scattering with final momentum $%
\mathbf{K}_{f}$, from a crystal surface in the initial state $\left\vert
\phi _{i}\right\rangle $, is obtained from Eq. (\ref{Apsivr}) as
\begin{equation}
\frac{dP_{i}^{{\small (\mathrm{P-SIVR})}}}{d\mathbf{K}_{f}}%
=\sum\limits_{f^{\prime }}\left\vert A\left[ a_{if^{\prime }}\right]
\right\vert ^{2},  \label{Pif}
\end{equation}%
where the sum over $f^{\prime }$ involves the addition of all the final
crystal states $\left\vert \phi _{f^{\prime }}\right\rangle $ satisfying the
total energy conservation.

In order to derive a more easy to handle expression for Eq. (\ref{Pif}), we
introduce a pairwise additive model to represent the projectile-surface
interaction. Within the pairwise model, $V_{PS}$ is built by adding the
binary interatomic potentials that describe the interaction of the atomic
projectile with individual ionic centers of the crystal. It reads

\begin{equation}
V_{PS}(\mathbf{R}_{t},\underline{\mathbf{u}})=\sum\limits_{\mathbf{r}_{%
\mathrm{B}}}v_{\mathbf{r}_{\mathrm{B}}}\left( \mathbf{R}_{t}-\mathbf{r}_{%
\mathrm{B}}-\mathbf{u}(\mathbf{r}_{\mathrm{B}})\right) ,  \label{Vps}
\end{equation}%
where $v_{\mathbf{r}_{_{\mathrm{B}}}}\left( \mathbf{r}\right) $ denotes the
binary projectile-ion interaction as a function of the relative vector $%
\mathbf{r}$, with $v_{\mathbf{r}_{\mathrm{B}}}=$ $v_{1}$ or $v_{2}$ to
consider the two different ions of the crystallographic basis. Replacing Eq.
(\ref{Vps})\ in Eq.(\ref{Vsp}), the crystal factor can be expressed as

\begin{eqnarray}
F_{if}^{(c)}(\mathbf{R}_{t},t) &=&\left( 2\pi \right) ^{-3/2}\sum\limits_{%
\mathbf{r}_{\mathrm{B}}}\int d\mathbf{q\ }\tilde{v}_{\mathbf{r}_{\mathrm{B}%
}}(\mathbf{q})e^{i\mathbf{q}\cdot (\mathbf{R}_{t}-\mathbf{r}_{\mathrm{B}})}
\notag \\
&&\times \left\langle \phi _{f}\left\vert \exp \left[ -i\mathbf{q}\cdot
\mathbf{U}_{t}(\mathbf{r}_{\mathrm{B}})\right] \right\vert \phi
_{i}\right\rangle ,  \label{vifq}
\end{eqnarray}%
where $\tilde{v}_{\mathbf{r}_{\mathrm{B}}}(\mathbf{q})$ is the Fourier
transform of $v_{\mathbf{r}_{_{\mathrm{B}}}}\left( \mathbf{r}\right) $ and $%
\mathbf{U}_{t}(\mathbf{r}_{\mathrm{B}})=\exp \left( iH_{S}t\right) \mathbf{u}%
(\mathbf{r}_{\mathrm{B}})\mathbf{\exp }\left( -iH_{S}t\right) $ is the
deviation operator within the Heisenberg picture \cite{Cohen}.

\medskip Finally, to compare with the experiments the differential
probability $dP_{i}^{{\small (\mathrm{P-SIVR})}}/d\mathbf{K}_{f}$, given by
Eq. (\ref{Pif}), must be averaged over the equilibrium distribution of the $%
\phi _{i}$- wave functions. Following a procedure similar to that given in
the Appendix N of Ref. \cite{Ashcroft}, after some steps of algebra that
involve the use of Eq. (\ref{vifq}), we obtain an averaged probability $dP^{%
{\small (\mathrm{P-SIVR})}}/d\mathbf{K}_{f}$, which includes a correlation
factor

\begin{eqnarray}
C(\mathbf{q},\mathbf{r}_{\mathrm{B}},t;\mathbf{q}^{\prime },\mathbf{r}_{%
\mathrm{B}}^{\prime },t^{\prime }) &=&\left\langle \exp \left[ i\mathbf{q}%
^{\prime }\cdot \mathbf{U}_{t^{\prime }}(\mathbf{r}_{\mathrm{B}}^{\prime })%
\right] \right.  \notag \\
&&\left. \exp \left[ -i\mathbf{q}\cdot \mathbf{U}_{t}(\mathbf{r}_{\mathrm{B}%
})\right] \right\rangle ,  \label{Caverage}
\end{eqnarray}%
where the averaged value $\ \left\langle X\right\rangle $ of any operator $X$%
, at the equilibrium temperature $T$, is given by Eq. (N.13) of Ref. \cite%
{Ashcroft}. The factor $C(\mathbf{q},\mathbf{r}_{\mathrm{B}},t;\mathbf{q}%
^{\prime },\mathbf{r}_{\mathrm{B}}^{\prime },t^{\prime })$ can be then
expanded as a power series%
\begin{eqnarray}
C(\mathbf{q},\mathbf{r}_{\mathrm{B}},t;\mathbf{q}^{\prime },\mathbf{r}_{%
\mathrm{B}}^{\prime },t^{\prime }) &=&\exp \left[ \mathbf{-}W_{\mathbf{r}_{%
\mathrm{B}}}\mathbf{(q)-}W_{\mathbf{r}_{\mathrm{B}}^{\prime }}\mathbf{(%
\mathbf{q}^{\prime })}\right]  \notag \\
&&\times \sum\limits_{n=0}^{+\infty }c_{n}(\mathbf{q},\mathbf{r}_{\mathrm{B}%
},t;\mathbf{q}^{\prime },\mathbf{r}_{\mathrm{B}}^{\prime },t^{\prime }),
\notag \\
&&  \label{cn}
\end{eqnarray}%
with
\begin{equation}
c_{n}(\mathbf{q},\mathbf{r}_{\mathrm{B}},t;\mathbf{q}^{\prime },\mathbf{r}_{%
\mathrm{B}}^{\prime },t^{\prime })=\frac{\left\langle \left[ \mathbf{q}%
^{\prime }\cdot \mathbf{U}_{t^{\prime }}(\mathbf{r}_{\mathrm{B}}^{\prime })%
\right] \left[ \mathbf{q}\cdot \mathbf{U}_{t}(\mathbf{r}_{\mathrm{B}})\right]
\right\rangle ^{n}}{n!},  \label{cnn}
\end{equation}%
and $W_{\mathbf{r}_{\mathrm{B}}}\mathbf{(q)}$ being the Debye-Waller
function, defined as
\begin{equation}
W_{\mathbf{r}_{\mathrm{B}}}\mathbf{(q)=}\frac{\left\langle \left[ \mathbf{q}%
\cdot \mathbf{u}(\mathbf{r}_{\mathrm{B}})\right] ^{2}\right\rangle }{2},
\label{DWfactor}
\end{equation}%
where the dependence on $\mathbf{r}_{\mathrm{B}}$ indicates that its value
changes for the different species of the crystallographic basis, as well as
for bulk or surface positions.

Using the expansion given by Eq. (\ref{cn}), the P-SIVR probability can be
expressed as a series

\begin{equation}
\frac{dP^{{\small (\mathrm{P-SIVR})}}}{d\mathbf{K}_{f}}=\sum\limits_{n=0}^{+%
\infty }\frac{dP_{n}}{d\mathbf{K}_{f}},  \label{Ptot}
\end{equation}%
where $dP_{n}/d\mathbf{K}_{f}$ accounts for the partial probability
corresponding to the $\mathbf{K}_{i}\rightarrow \mathbf{K}_{f}$ \ transition
with the exchange of $n$ phonons. It reads

\begin{eqnarray}
\frac{dP_{n}}{d\mathbf{K}_{f}} &=&\sum\limits_{\mathbf{r}_{\mathrm{B}},%
\mathbf{r}_{\mathrm{B}}^{\prime }}\int d\mathbf{q}\int d\mathbf{q}^{\prime
}\int\limits_{0}^{+\infty }dt\int\limits_{0}^{+\infty }dt^{\prime }\ c_{n}(%
\mathbf{q},\mathbf{r}_{\mathrm{B}},t;\mathbf{q}^{\prime },\mathbf{r}_{%
\mathrm{B}}^{\prime },t^{\prime })  \notag \\
&&\times A[b_{t}(\mathbf{q,r}_{\mathrm{B}})]\ A[b_{t^{\prime }}^{\ast }(%
\mathbf{q}^{\prime }\mathbf{,r}_{\mathrm{B}}^{\prime })],  \label{Prob-n}
\end{eqnarray}%
with $A[b]$ defined by Eq. (\ref{Apsivr}) and%
\begin{eqnarray}
b_{t}(\mathbf{q,r}_{\mathrm{B}}) &=&\ \left\vert J_{P}(t)\right\vert
^{1/2}e^{i\nu _{t}\pi /2}\ \tilde{v}_{\mathbf{r}_{\mathrm{B}}}(\mathbf{q}%
)\exp \left[ \mathbf{-}W_{\mathbf{r}_{\mathrm{B}}}\mathbf{(q)}\right]  \notag
\\
&&\times \exp \left[ i\left( \varphi _{t}-\mathbf{Q}\cdot \mathbf{R}_{o}+%
\mathbf{q}\cdot (\mathbf{R}_{t}-\mathbf{r}_{\mathrm{B}})\right) \right] .
\notag \\
&&  \label{ampli-aver}
\end{eqnarray}%
From Eq. (\ref{Prob-n}) we derive more compact expressions for the orders $%
n=0$ and $n=1$, corresponding to the partial probabilities for zero- and
one-phonon scattering, which are given in the text by Eqs. (\ref%
{prob-P0-sivr}) and (\ref{prob-P1-sivr}), respectively.

\bibliographystyle{unsrt}
\bibliography{fononLiF}

\end{document}